\documentclass[ijds,nonblindrev]{informs5}
\DoubleSpacedXI

\usepackage{amsmath,amssymb,amsfonts}

\usepackage{natbib}
 \bibpunct[, ]{(}{)}{,}{a}{}{,}%
 \def\bibfont{\small}% 

\usepackage{rotating}
\usepackage{fancyvrb}

\usepackage[ruled]{algorithm2e}
\usepackage{appendix} 

\TheoremsNumberedThrough     % Preferred (Theorem 1, Lemma 1, Theorem 2)
\ECRepeatTheorems
\JOURNAL{}

\EquationsNumberedThrough    % Default: (1), (2), ...
\MANUSCRIPTNO{IJDS-0001-1922.65}

\begin{document}
%\setcounter{page}{1}
%\DOI{10.1287/xxxx.0000.0000}%


\RUNTITLE{Data-driven Option Pricing}

\TITLE{Data-driven Option Pricing\footnote{Dai acknowledges the supports of 
		Hong Kong GRF (15217123, 15213422), The Hong Kong Polytechnic University Research Grants (P0039114, P0042456,
		P0042708, and P0045342), and NSFC (12071333).}}

\ARTICLEAUTHORS{%
\AUTHOR{Min Dai\textsuperscript{a}, Hanqing Jin\textsuperscript{b}, Xi Yang\textsuperscript{c}}

\AFF{\textsuperscript{a}Department of Applied Mathematics and School of Accounting and Finance, The Hong Kong Polytechnic University, Kowloon, Hong Kong, \EMAIL{mindai@polyu.edu.hk};
\textsuperscript{b}Mathematical Institute, University of Oxford, Andrew Wiles Building, Radcliffe Observatory Quarter, Woodstock Road, Oxford, OX2 6GG, United Kingdom, \EMAIL{jinh@maths.ox.ac.uk}; 
\textsuperscript{c}Department of Mathematics, National University of Singapore, Block S17,10 Lower Kent Ridge Road, Singapore, \EMAIL{yangxi@u.nus.edu}}

%mirko.janc@informs.org
%\AUTHOR{}
%
%\AFF{}
%
%\AUTHOR{}
%
%\AFF{}
}

\ABSTRACT{%
We propose an innovative data-driven option pricing methodology that relies exclusively on the dataset of historical underlying asset prices. While the dataset is rooted in the objective world, option prices are commonly expressed as discounted expectations of their terminal payoffs in a risk-neutral world. Bridging this gap motivates us to identify a pricing kernel process, transforming option pricing into evaluating expectations in the objective world. %We reveal that the pricing kernel process corresponds to the reciprocal of the optimal wealth process linked to a dynamic logarithmic utility optimization problem and that estimating the expectations can be reduced to a functional optimization problem. 
We recover the pricing kernel by solving a utility maximization problem, and evaluate the expectations in terms of a functional optimization problem. %apply it in the pricing for any financial derivative, for which the conditional  expectations can be formulated as an optimisation over pricing functions.
Leveraging the deep learning technique, we design data-driven algorithms to solve both optimization problems over the dataset. Numerical experiments are presented to demonstrate the efficiency of our methodology.
}

%\SUBJECTCLASS{}
%\AREAOFREVIEW{}

\maketitle

%\newpage

\section{Introduction}\label{sec1}
\citet{Black1973} develop the celebrated Black-Scholes option pricing model. However, the constant volatility assumed in the model contradicts the volatility smile phenomenon prevalent in options markets.\footnote{See, e.g., \cite{Hull2014}.} To align with the volatility smile phenomenon, existing literature proposes various advanced option pricing models beyond the Black-Scholes world, such as local volatility models, stochastic volatility models, and jump-diffusion models.\footnote{See, e.g., \cite{Dupire1994}, \cite{Heston1993, Hull1987}, \cite{Merton1976}, and \cite{Kou2002}.} Calibrating these models often requires an ample and reliable dataset of (vanilla) option prices, which is, however, absent in many emerging options markets. For example, there are no exchange-traded options on individual stocks in China. Even in the U.S., only a limited number of exchange-traded options are written on less liquid stocks. 

This paper aims to develop a novel data-driven option pricing methodology exclusively reliant on historical underlying asset prices. %In other words, we attempt to recover option prices solely from past underlying asset prices. 
This pricing methodology dispenses with a routine calibration step that involves market prices of options, and as a consequence, it can apply to derivatives pricing in emerging derivatives markets. Our methodology shares similarities with the deep hedging method \citep[e.g.,][]{Buehler2019}, both utilizing the deep learning technique and historical underlying asset prices. However, unlike the deep hedging method, which finds risk-indifference prices of a derivative, our methodology instead finds a pricing kernel that links the objective probability measure to a risk-neutral probability measure. %The pricing kernel allows us to utilize our dataset for option pricing. 
The pricing kernel enables us to price any derivatives on the (same) underlying asset by evaluating an expectation in the objective probability measure via historical underlying asset prices.

Next we elaborate on our data-driven pricing methodology. For illustration, we assume that our data is the historical underlying asset prices generated by an Ito process, where the drift and volatility terms are {\it unknown} deterministic functions of the underlying asset price. Crucially, these terms are presumed to be time-independent as our methodology attempts to extract the price information of options solely from our dataset. 

While our dataset is rooted in the objective world, option prices are commonly expressed as discounted expectations of their terminal payoffs in the risk-neutral world. Bridging this gap via a data-driven method motivates us to identify a pricing kernel process, transforming option pricing into estimating expectations in the objective world. Our key idea is to introduce a dynamic logarithmic utility optimization problem whose optimal wealth process turns out to be the reciprocal of the pricing kernel process. The utility optimization problem naturally induces a loss function, which can be evaluated over the sample paths drawn from our dataset. Minimizing the loss function leads to the optimal wealth process, or equivalently, the pricing kernel process.  

Once the pricing kernel process is available, pricing a derivative is then reduced to estimating the expectation of its terminal payoff multiplied by the pricing kernel in the objective world. Inspired by \citet{Jia2022}, we estimate the expectation by minimizing a {\it martingale loss function}, which is also evaluated over our dataset.

%solving a functional optimization problem with deep learning. 

Leveraging the deep learning technique, we design data-driven algorithms to minimize the loss functions over our dataset. As a learning algorithm usually requires a huge amount of data and our method uses historical data of underlying asset prices rather than simulated data, we have to allow overlapping sample paths. As such, the initial prices of different sample paths differ in general. Assuming time-homogeneity of market parameters, we rephrase our optimization problems to exclude the dependence on the initial prices of sample paths, and the minimizers we try to find are functions of the underlying asset price. We present numerical experiments to demonstrate the efficiency of our algorithms. 
%	{\color{red}	We need to emphasis that our pricing in this paper only use historical data of underlying asset prices. This setting is a big challenging for data collecting, since learning algorithm usually need huge amount of data, while we cannot go to very deep history to keep the relevance of the data to the market in the near future.   }

\paragraph{Related Literature.} 

Recently, data-driven option pricing methods have attracted much research interest. Instead of specifying a particular underlying market model, these methods make mild assumptions about the market and directly utilize the dataset of historical prices. One example is a deep hedging method proposed by \citet{Buehler2019}, who utilize a deep  learning technique for risk-indifference pricing and hedging of derivatives in frictional markets. Our approach differs significantly from this method in two primary aspects. First, our approach builds on the martingale theory of derivatives pricing. As such, the derivative prices obtained from our approach are theoretically arbitrage-free and interpretable. In contrast, arbitrage opportunities may exist among derivative prices derived from deep hedging. Second, the deep hedging method necessitates separate training for different derivatives on the same underlying. In contrast, our approach conducts a single training phase to estimate the pricing kernel that applies to all derivatives on the same underlying, thereby reducing the amount of computation significantly. 
%Notably, while our approach is tailored to complete markets, it retains the ability to generate arbitrage-free option prices in incomplete market scenarios.

To construct the pricing kernel, we introduce a dynamic portfolio optimization problem. %which can be solved by a data-driven approach combined with deep learning. 
There is rich literature on data-driven portfolio optimization with reinforcement learning. For example, \citet{Wang2020a} propose a general continuous-time exploratory stochastic control framework with reinforcement learning, where an entropy term is introduced to encourage exploration. Their framework can be applied to many portfolio optimization problems; see, e.g., \citet{Wang2020b}, \citet{Guo2020}, and \citet{Dai2023}.  \citet{Hambly2021} provide a comprehensive review of recent advancements in the application of reinforcement learning in finance, including portfolio optimization.   
%delve into the implications of entropy regularization within the context of mean field games (MFGs) with learning over a finite time horizon. \citet{Gu2020} introduce a machine learning approach by employing the autoencoders technique.
While it is promising to solve our dynamic portfolio optimization problem with reinforcement learning, we adopt a deep learning approach over the dataset, where the policy function is parameterized by neural networks. %and a.% and solve the problem using a stochastic gradient descent algorithm.

\citet{Jia2022} show that policy evaluation (PE) in continuous-time reinforcement learning is identical to maintaining the martingale condition. They then design PE algorithms based on the martingale characterization. As evaluating expectations involved in the second phase of our method is analogous to PE, we minimize the {\it martingale loss function} proposed in \citet{Jia2022} to obtain the pricing functions of derivatives.

% With the pricing kernel obtained in the first phase, the evaluation of options  turns out to be the evaluation of a (conditional) expectation (under the real world probability measure).   \citet{Jia2022} proposed a smart data-driven approach to evaluate conditional expectations by martingale property, which inspired us to evaluate the price of our derivatives by the same way. 

{

	 The remainder of the paper is organized as follows. In Section 2, we first describe the market data. Then we present our methodology, where a critical step is to introduce a dynamic optimization problem to find a pricing kernel linking the risk-neutral probability measure to the objective measure. In Section 3, we design data-driven algorithms to find the price kernel and evaluate expectations in the objective world based on our dataset. We present our numerical experiments in Section 4 and potential future research directions along this line in Section 5. All technical proofs are relegated in Appendix.
}

\section{Market Data and Methodology}\label{sec2}
\paragraph{Market Data:} 
Consider a continuous-time financial market consisting of a risk-free asset (bond) and a risky asset (stock).\footnote{%Although this paper restricts attention to the pricing with a single underlying asset, 
	It is straightforward to extend our methodology to a complete market with multiple underlying assets.}
	 The stock price process is described by % follows a continuous It\^{o} process, i.e., it satisfy 
the following stochastic differential equation (SDE): %in the form
\begin{equation}
	\label{underlying_process}
	\frac{dS_t}{S_t} = \mu(S_t)dt + \sigma(S_t) dB_t,\ \ \ S_0 = s_0 \in \mathbb{R}^{+}, 
\end{equation}
where $B=\{B_t, t\geq 0\}$ is a standard Brownian motion on filtered probability space $(\Omega, \mathcal{F},\mathbb{P};\{\mathcal{F}_t\}_{t\geq 0})$, and the drift term $\mu(\cdot): \mathbb{R} \rightarrow \mathbb{R}$ and the volatility term $\sigma(\cdot): \mathbb{R}\rightarrow\mathbb{R}$ are some deterministic (but unknown) functions. Assume that the risk-free interest rate $r$ is a known constant,\footnote{We can also assume  that $r$ is a known deterministic function of $t$.} 
and the historical stock prices $S_{\tau},$ $\tau\leq t$, are observable at time $t$. It is worth highlighting that the assumption of %We assume that $\mu$ and $\sigma$ are independent of $t$ as  
time-homogenous dynamics of the stock price process enables us to utilize historical stock data to unveil option prices. %This is quite a critical assumption, without which we cannot take advantage of historical data for the study of the market in the future. In addition, 

Technically, we need the following standard regularity assumptions.

\begin{assumption} \label{assumption2}
		The following conditions hold true:
\begin{itemize}
	\item [(i)] The SDE \eqref{underlying_process} admits a unique strong solution.\footnote{This condition can be weakened into the existence and uniqueness of a weak solution.}
	\item [(ii)] The  market admits at least one equivalent martingale measure $\mathbb{Q}$ (known as a risk-neutral measure), which implies that the market  is arbitrage-free. %, i.e., there exist at least an equivalent martingale measure $\mathbb{Q}$;
	\item [(iii)] The equivalent martingale measure $\mathbb{Q}$ is unique. %, which implies that the equivalent martingale measure is unique.  
\end{itemize}
\end{assumption}

{Parts (i) and (ii) in the above assumption are mild and necessary for derivatives pricing. %  is clearly very mild and  quite necessary for a market model to be reasonable, with which $S_t$ is almost surely strictly positive; the arbitrage-free property in (ii) is also necessary for any pricing to be meaningful; 
	While part (iii) may not be necessary, we include it to ease exposition. %in order to simplify the presentation of our idea. 
	If part (iii) does not hold, then a financial derivative may permit multiple no-arbitrage prices, and the price obtained in this paper will be merely one of the no-arbitrage prices.\footnote{In this case, one may try a class of utility functions, rather the logarithm utility function only, to obtain an interval of the no-arbitrage prices. We leave it for future research.}}

%In Assumption  \ref{assumption2}, (i)  is clearly very mild and  quite necessary for a market model to be reasonable, with which $S_t$ is almost surely strictly positive; the arbitrage-free property in (ii) is also necessary for any pricing to be meaningful; (iii) is not necessary, but if (iii) does not hold, then the no-arbitrage price of a financial derivative may not be unique. 

Consider a contingent claim (option) with payoff $\xi_T$ at maturity $T$, where $\xi_T$ is an $\mathcal{F}_T$-measurable random variable (with a certain regularity). Under Assumption \ref{assumption2}, we can express the option price as
%\begin{equation}
%	\label{Q_pricing_formula0}
%	V_t= \mathbb{E}^\mathbb{Q}\left[e^{-\int_t^Tr_udu}\xi_T\mid \mathcal{F}_t\right], \quad 0\leq t \leq T.
%\end{equation}
\begin{equation}
	\label{Q_pricing_formula0}
	V_t= \mathbb{E}^\mathbb{Q}\left[e^{-r(T-t)}\xi_T\mid \mathcal{F}_t\right], \quad 0\leq t \leq T.
\end{equation}
%where $s$ is the underlying asset price at time $t$. 
% namely, %only depends on the time to maturity and the spot price $S_t$, hence we have the pricing function $V(t,s)$ such that 
%\begin{equation}
%	\label{Q_pricing_formula}
%	V_t= V(t, S_t)=\mathbb{E}^\mathbb{Q}\left[e^{-\int_t^Tr_udu}h(S_T)\mid \mathcal{F}_t\right], \quad 0\leq t \leq T.
%\end{equation}

If  $\sigma(\cdot)$ is known, then the option price could be evaluated by Monte-Carlo simulations in the risk-neutral world or numerical methods for partial differential equations (PDEs). %In a model-based pricing, $\sigma$ should be firstly determined. 
To identify $\sigma(\cdot)$, a popular model-based method is based on the Dupire equation, requiring the availability of market prices of vanilla options with (as many as possible) different strikes and expiries
\citep[e.g.,][]{Bouchouev1997, Jiang2003}. In this paper, we assume insufficient option data and aim to propose a stock-data-driven option pricing algorithm where we do not estimate $\sigma(\cdot)$. %in advance. %Unfortunately, the lack of enough option prices often happens in reality.  
 %The prerequisite for it  is local volatility calibration to determine parameter function $\sigma$. Usually, calibration requires market price data of options with high liquidity. This condition cannot be fulfilled in many cases. 

\paragraph{Methodology:}
%In this paper, we propose a novel option pricing algorithm that only requires historical price data of the underlying asset without the estimation of $\sigma$. It is applicable in a wider range of scenarios compared to model-based methods.

Under Assumption \ref{assumption2}, we define the Radon-Nikodym derivative of the equivalent martingale measure $\eta_t = \frac{d\mathbb{Q}}{d\mathbb{P}}\mid_{\mathcal{F}_t}$ and the pricing kernel 
%\begin{equation}
%	\label{pricing_kernel}
%	\rho_t = \eta_t e^{-\int_0^tr_sds}.
%\end{equation}
\begin{equation}
	\label{pricing_kernel}
	\rho_t = \eta_t e^{-rt}.
\end{equation}
Then we can rewrite equation \eqref{Q_pricing_formula0} as %any (lower bounded) payoff $\xi_T$ paid at maturity $T$ should be priced at $t$ by
\begin{equation}
	\label{price_formual}
	V_t=\frac{1}{\rho_t} \mathbb{E}[\xi_T\rho_T\mid \mathcal{F}_t],
\end{equation}
where $\mathbb{E}$ refers to the expectation under the objective probability measure $\mathbb{P}$. 
Using the above pricing formula, we can price the option by evaluating the expectation under the probability measure $\mathbb{P}$ provided that the pricing kernel $\rho$ and sufficient historical stock prices are available.

To construct the pricing kernel $\rho$, we consider a %from the perspective of
 logarithmic utility maximization problem. Assume that an investor with a (unit) initial wealth of $X_0 = 1$ trades in the market. % to maximize the terminal wealth. 
 The investor's self-financing wealth process $X_t$ satisfies:
\begin{equation}
	\label{portfolio_process}
	d X_t=r (X_t-\pi_t) d t+ \pi_t\frac{d S_t}{S_t} ,
\end{equation}
where $\pi_t $ is the dollar amount invested in the stock.
The investor aims to maximize the expected  log return of the investment over the horizon $[0,T]$ by choosing an admissible strategy, namely,
\begin{equation}
	\label{log_util}
	\max_{\pi\in\mathcal{A}}\mathbb{E}\left[\operatorname{ln}\left(X_T\right)\right],
\end{equation}
where $\mathcal{A}$ is the set of admissible strategies starting from the initial wealth $X_0=1$. %, and $T$ is the investment horizon. 

\iffalse
If the risky asset price process is geometric Brownian Motion (GBM), this stochastic control problem is known as Merton Problem \citeyearpar{merton_problem}, which could be solved by the Hamilton-Jacobi-Bellman (HJB) equation, or the martingale method \citep{COX}. For the general drift term $\mu(\cdot)$ and volatility term $\sigma(\cdot)$ in our paper, the two methods described above can be employed to solve problem \eqref{log_util}. In Theorem \ref{theorem_log_util}, it not only provides the optimal portfolio for the logarithmic utility maximization problem but also reveals the relationship between optimal wealth $X_t$ and the pricing kernel $\rho_t$.
\fi
If $\mu$ and $\sigma$ are known, problem \eqref{log_util} can be solved explicitly, as shown in the following theorem.
\begin{theorem}
	\label{theorem_log_util} Assume that $\mu(\cdot)$ and $\sigma(\cdot)$ are known.
	The optimal portfolio for problem \eqref{log_util} is given by
	$$
	\pi_t^*=X_t\frac{\mu(S_t)-r}{\sigma^2(S_t)},
	$$
and the optimal wealth process is given by
	$$
	X_t=\rho_t^{-1}.
	$$
\end{theorem}

The proof of Theorem \ref{theorem_log_util} is relegated to Appendix.
\iffalse
 This theorem gives us the optimal investment for logarithmic utility function if we have the drift and volatility functions, which we don’t really know.
 \fi
 Although we do not know $\mu$ and $\sigma$, Theorem \ref{theorem_log_util} inspires us to solve problem  \eqref{log_util} by a data-driven method, which we elaborate on next section, to recover the pricing kernel process $\rho_t$.   

%In the next section, we aim to use historical data to construct the pricing kernel from logarithmic utility maximization problem perspective and evaluate conditional expectation (\ref{price_formual}).

\section{Algorithm Design}\label{sec3}

In this section, we design data-driven algorithms to recover the pricing kernel process $\rho_t$ by solving problem \eqref{log_util} % logarithmic utility maximization problem
 and evaluate the conditional expectation in the pricing formula \eqref{Q_pricing_formula0}.
 % from the perspective of functional optimization problem. 
 The dataset required by our algorithms is the historical stock prices only. % of the underlying asset  in a fixed-length time interval $[t_i,t_i +T]$. Those time window can overlap each other if there are no enough data.

\subsection{Recovery of the Pricing Kernel $\rho_t$}\label{pricingkernel}
By Theorem \ref{theorem_log_util}, the optimal strategy for problem \eqref{log_util} must be in the form of $\pi^*_t = X_tf^*(S_t)$. %with some deterministic function $f: \mathbb{R}^+\rightarrow \mathbb{R}$ to be determined. 
 Therefore, we will maximize  $\mathbb{E}\left[\operatorname{ln}\left(X_T\right)\right]$ among all portfolios in the form of $\pi_t/X_t = f(S_t)$, namely,
\begin{eqnarray}\label{optf}
	\max_{f} \hat{L}(f\mid s_0),
\end{eqnarray} %whose objective value is  %  Applying It\^{o}'s lemma for formula (\ref{portfolio_process}) with $\pi_t = X_t f(S_t)$,  we have
where
\begin{eqnarray}
	\label{log_util_s}
\hat{L}(f\mid s_0)&=:&	\mathbb{E}\left[\ln(X_T)\mid X_0 = 1, S_0 = s_0\right] \nonumber\\
 & = & \mathbb{E}\left[\int_0^Tr(1-f(S_t))dt + \frac{f(S_t)}{S_t}dS_t - \frac{f^2(S_t)}{2S^2_t}d\langle S,S\rangle_t\mid S_0 =s_0\right].
	%&=:& \hat{L}(f\mid s_0).
\end{eqnarray}

We plan to employ a deep learning algorithm to solve problem \eqref{optf}, where we use sample paths drawn from the dataset to estimate the expectation in \eqref{log_util_s} for any given function $f$.
It is worth pointing out that the sample paths required in this estimation  should share the same initial stock price $s_0$. However, different sample paths in our dataset usually start from different initial prices. Hence, what we can calculate by our dataset is an estimate of 
\begin{equation}
	\label{loss_f_dl_diff}
	L(f) = \mathbb{E}_{s_0\sim \omega}\left[\mathbb{E}\left[\int_0^Tr(1-f(S_t))dt + \frac{f(S_t)}{S_t}dS_t - \frac{f^2(S_t)}{2S^2_t}d\langle S,S\rangle_t\mid S_0 =s_0\right]\right],
\end{equation}
where $\omega$ represents the distribution followed by the initial prices of the sample paths in our dataset. Fortunately, the following proposition helps overcome the obstacle.  %function $L(f)$ and $\hat{L}(f \mid s_0)$ share the same  maximizer $f^*$, which is reformulated in proposition \ref{prop_loss_same_min}.
\begin{proposition}
	\label{prop_loss_same_min}
	%In the logarithmic utility maximization problem, 
	Function $\hat{L}(\cdot \mid s_0)$ with any $s_0\in \mathbb{R}^+$ and function $L(\cdot)$ share the same and unique maximizer $f^*$. %, which is also the  unique maximizer of  $L(f)$. 
\end{proposition}

We relegate the proof for this proposition to Appendix. %\ref{proof_prop_loss_same_min}. 

Thanks to Proposition \ref{prop_loss_same_min}, we can replace $\hat{L}(f\mid s_0)$ with $L(f)$ when solving problem \eqref{optf}. Below we elaborate on our data-driven algorithm for problem \eqref{optf}. %we can obtain the optimal strategy of the original problem ($\ref{log_util}$) by maximizing $L(f)$ in terms of a large number of sample paths with different initial prices. %empirical distribution $\omega$ of the initial price $s_0$. Given enough historical stock price data, we can use a deep learning method to solve the maximization problem. 
We parameterize function $f(\cdot)$ by a multilayer feedforward neural network, % taking underlying asset price as inputs, 
that is, {$f(s ; \phi) = NN^\phi(s)$}. 
{%Let $S_{t_i}$ be the (historical) stock price at time $t_i=i\Delta t$, where $\Delta t$ is the step size in time.
%We sample the historical stock prices in a discrete-time setting with a fixed time step $\Delta t$. 
Suppose we sample $J$ paths of the stock price from the dataset, and each path consists of $N+1$ consecutive stock prices with the time step size of $\Delta t$. Denote by
$S_{n,j}$ the stock price at the $n$-th time spot in the $j$-th sample path, where $j = 0,1,\cdots, J-1$, $n=0,1,\cdots, N$, and $T=N\Delta t$. With these notations,  $\left\{S_{0,j},S_{1,j},\cdots, S_{N,j}\right\}$ represents the $j$-th sample path. Note that these paths sampled from the dataset may overlap each other to make more samples for our learning algorithm.

 For any function $f(\cdot; \phi)=NN^\phi(\cdot)$, we define the following loss function $l(\phi)$, which is a discrete-time version of $-L(f)$:   
 % which is the discrete form of $-L(f)$ after parameterizing the function $f$ with $\phi$. Since loss function typically refers to a minimization problem, we add a negative sign in front of the expression.
\begin{equation}
	\label{loss_f_dl_diff_discrete}
	l(\phi) = -\frac{1}{J}\sum_{j=0}^{J-1} l_j(\phi),
\end{equation}
where 
\begin{equation}
l_j(\phi)=	\sum_{n=0}^{N-1}\left(r(1-f(S_{n,j};\phi))\Delta t + \frac{f(S_{n,j};\phi)(S_{{n+1,j}}-S_{n,j})}{S_{n,j}} - \frac{f^2(S_{n,j};\phi)(S_{{n+1,j}}-S_{n,j})^2}{2S_{n,j}^2}\right).
\end{equation}

Note that the loss function does not rely on the knowledge of the drift function $\mu$ and the diffusion function $\sigma$. Using the dataset, we can implement the stochastic gradient descent algorithm \citep[e.g.,][]{Adam2014} to minimize the loss function $l(\phi)$ and obtain the minimizer $\phi^*$ .
} 
%The updating rule for parameters $ \phi $ will be provided in Algorithm %\ref{alg:data_driven_option_pricing}.

Once function {$f^*(\cdot) = f(\cdot;\phi^*)$} is available, we can use \eqref{portfolio_process} to construct the pricing kernel $\rho_t$ for any stock price path.
\subsection{Estimation of Conditional Expectation}\label{est_cond_exp}

\iffalse
If the underlying asset price paths from time $t$ to maturity $T$ with the fixed price $S_t$ are available, the corresponding pricing kernel paths could be calculated by estimated function $f$. And the option price $V(t,S_t)$ at time $t$ with spot price $S_t$, an expectation of random variable $\frac{\xi_T\rho_T}{\rho_t}$, could be approximated by the mean of samples drawn from $\frac{\xi_T\rho_T}{\rho_t}$. In practice, there is only one price path available, i.e., historical price data. Hence, it is impractical to use the above method to calculate option value directly. While we collect the historical prices of underlying asset over time windows of $(T-t)$-length as samples in our dataset, different samples may possess different initial value $S_0$. It is not justifiable to employ the mean of the product of the payoff and the terminal values of the pricing kernel process as the option price.
\fi

In this section, we will evaluate the option price $V_t$ through \eqref{price_formual} in terms of the pricing kernel $\rho_t(\cdot)$ obtained. %for any historical price path. 
%consider the pricing of a path independent option with payoff $\xi_T=h(S_T)$, whose price at time $t$ can be determined by a pricing function $V(t, S_t)$.
If there are sufficient sample paths of stock price starting from a fixed initial price $S_t=s$, %then for a path independent option $\xi_T$, 
we can first construct $\rho_T/\rho_t$ for each sample path and then estimate the option price $V_t$ by the sample mean of $\xi_T\rho_T/\rho_t$. However, in practice, there are either no or a very limited number of such sample paths starting from the same stock price $s$. As such, we have to use a different approach.
%and the operation above is  impractical. Even if we use the sliding window to sample different paths of the underlying asset prices, the initial prices will be typically different over different time window, and the sample mean of $\xi_T\rho_T/\rho_t$ from different paths is meaningless. 

For illustration, we assume that the option's payoff is path-independent,\footnote{It is straightforward to extend our pricing idea to path-dependent options.} i.e., $\xi_T=h(S_T)$. Then we can write $V_t=V(t,S_t)$, which is a conditional expectation given $(t,S_t)$. We plan to learn the price function $V(\cdot,\cdot)$ using the dataset of the historical stock prices.

Inspired by \citet{Jia2022}, we transform the estimation for conditional expectations into a functional optimization problem. %A fundamental property of conditional expectations yields that 
{Indeed, since $\rho_t V(t,S_t)=\mathbb{E}[{\xi_T\rho_T}\mid \mathcal{F}_t]$ is a martingale under probability measure $\mathbb P$,} the option price process $V(t,S_t)$ minimizes the $L^2$-error between $\frac{\xi_T\rho_T}{\rho_t}$ and any $\mathcal{F}_t$-measurable random variables. It follows
\begin{equation}
	\label{conditional_expect}
	V(t, s) = \argmin_{y\in\mathbb{R}}\mathbb{E}\big[|y-\frac{\xi_T\rho_T}{\rho_t}|^2\ \big | \mathcal{F}_t,\  S_t=s\big].
\end{equation}
%In particular, the option price with spot price $s_0$ at time $t=0$ is given by
%\begin{equation}
%	\label{conditional_expect_s0}
%	V_0(s_0) = V(0, s_0) = \argmin_{y\in\mathbb{R}}\mathbb{E}\big[|y-\xi_T\rho_T|^2 \mid  S_0=s_0\big],
%\end{equation}
In particular, we restrict attention to the price function at time $t=0$, denoted by $V_0(s)$. 
Noticing $\rho_0=1$ and using a similar argument as in the proof of Proposition \ref{prop_loss_same_min}, we obtain 
\begin{equation}
	\label{conditional_expect_diff}
			V_0(\cdot) = \argmin_{\zeta\in\mathcal{B}(\mathbb{R})}\mathbb{E}_{S_0\sim \omega}\left[\mathbb{E}\big[|\zeta(S_0) -\xi_T\rho_T|^2 \ \big| S_0  \big]\right].
\end{equation}
Solving the functional minimization problem \eqref{conditional_expect_diff} allows us to use the sample paths with different initial prices and is thus feasible with our dataset.

{We parameterize $V_0(\cdot)$ by a multilayer feedforward neural network, that is, $V_0(s;\varphi) = NN^\varphi(s)$. For the $j$-th sample path, we denote \begin{equation}
	\label{loss_varphi_diff}
	Q_j(\varphi) = \left(V_0(S_{0,j};\varphi)-\xi_{N,j}\rho_{N,j}\right)^2,
\end{equation}
where $\xi_{N,j} = h(S_{N,j})$, and $\rho_{N,j}$ is the terminal value of the pricing kernel process associated with the $j$-th sample path.
Owing to \eqref{conditional_expect_diff}, we define the following loss function $Q(\varphi)$:
\begin{equation}
	\label{loss_varphi_diff0}
	 Q(\varphi) = \frac{1}{J}\sum_{j=0}^{J-1} Q_j(\varphi).
\end{equation}
Using the dataset, we can apply the stochastic gradient descent algorithm to minimize the loss function $Q(\varphi)$ and obtain the minimizer $\varphi^*$. Then $V_0(\cdot; \varphi^*)$ is the estimated option price function with time to maturity $T$.

We summarize the above data-driven option pricing method as pseudo-codes presented in Algorithm \ref{alg:data_driven_option_pricing},  where we adopt a single training sample to compute gradients when implementing the stochastic gradient descent algorithm to minimize the loss functions given in \eqref{loss_f_dl_diff_discrete} and	\eqref{loss_varphi_diff0}. In our numerical experiments, we use a mini-batch stochastic gradient descent algorithm. 
\SingleSpacedXI
\begin{algorithm}
	\caption{Pseudo-Codes for Data-driven Option Pricing}
	\label{alg:data_driven_option_pricing}
	\hspace{-0.2cm}\textbf{Input:} Horizon $T$, time step $\Delta t$, historical underlying asset prices $\{S_{n, j}: n=0,\cdots, N\}_{j=0}^{J-1}$, riskless interest rate $r$, number of time grids $N$, number of episodes $J$, the parametrized functions $f(s; \phi)$ and $V_0(s; \varphi)$,
	initial learning rates $\alpha_\phi$, $\alpha_\varphi$,  and learning rate schedule function $\gamma(\cdot)$ (a function of the number of episodes).\\
	\hspace{-0.2cm}\textbf{Learning procedure:}\\
	Initialize $\phi$, $\varphi$.\\
	\textbf{Learning optimal policy $f(\cdot;\phi^*)$}\\
	\For{episode $j=0$ to $J-1$}
	{
		Compute loss function
		$$ l_j(\phi) = -\sum_{n=0}^{N-1}\left(r(1-f(S_{n,j};\phi))\Delta t + \frac{f(S_{n,j};\phi)(S_{n+1,j}-S_{n,j})}{S_{n,j}} - \frac{f^2(S_{n,j};\phi)(S_{n+1,j}-S_{n,j})^2}{2S_{n,j}^2}\right);$$
		Update $\phi$ by 
		$
		\phi\leftarrow\phi -\gamma(j)\alpha_\phi\frac{\partial l_j(\phi)}{\partial \phi}
		$
	}
	\textbf{Learning the option value function $V_0(\cdot; \varphi^*)$}\\
	\For{episode $j=0$ to $J-1$}
	{
		Initialize $n =0$, $X_{0,j} = 1$\;
		\While{$n<N$}
		{
			Compute $X_{{n+1,j}} = X_{n, j}  [1+r\Delta t + f(S_{n,j};\phi)((S_{{n+1},j}-S_{n,j})/S_{n,j}-r\Delta t)]$ \;
		}
	Compute $\rho_{{N},j} = X_{{N}, j}^{-1}$;\\
		Compute terminal payoff $\xi_{N,j} = h(S_{N,j})$\;
		Compute 
		$$
		\Delta \varphi = \left(V_0(S_{0,j};\varphi)-\xi_{N,j}\rho_{N,j}\right)\frac{\partial V_0(S_{0,j};\varphi)}{\partial \varphi};
		$$
		Update $\varphi$ by 
		$
		\varphi\leftarrow\varphi - \gamma(j)\alpha_\varphi\Delta \varphi
		$
	}
\end{algorithm}
\DoubleSpacedXI
It is worth noting that we can estimate the option price function $V(t, s)$ simultaneously for all $t<T$. To do this, we can parameterize $V(t, s)$ by $V(t, s; \psi)=NN^{\psi}(s,t)$. %Similar to the martingale loss introduced in Jia and Zhou  \citeyearpar{jia_zhou_EP}, the loss function inspired 
Owing to \eqref{conditional_expect}, we can introduce the following martingale loss function as in \citet{Jia2022}:
\begin{equation}
	\label{rho_g_opti_pro}
	\text{ML}(\psi):=\mathbb{E}\left[\int_0^T\left|V(t,S_t;\psi)-\frac{\xi_T\rho_T}{\rho_t}\right|^2\mathrm{d}t\right],
\end{equation}
whose discrete-time form is 
%\begin{equation}
%	\label{rho_g_dicrete}
%	\min_{\psi}\text{ML}_{\Delta t}(\psi):= \frac{1}{2}\mathbb{E}\left[\sum_{i=0}^{N-1}\left(V(t_{i},S_{t_i};\psi)-\frac{\xi_T\rho_T}{\rho_{t_i}}\right)^2\Delta t\right].
%\end{equation}
\begin{equation}
	\label{rho_g_dicrete}
	\text{ML}_{\Delta t}(\psi):= \frac{1}{J}\sum_{j=0}^{J-1}\sum_{n=0}^{N-1}\left(V(n\Delta t,S_{n,j};\psi)-\frac{\xi_{N,j}\rho_{N,j}}{\rho_{n,j}}\right)^2\Delta t.
\end{equation}
%It is proved in Jia and Zhou  \citeyearpar{jia_zhou_EP} that minimizing loss function $\text{ML}(\psi)$ is equivalent to minimizing the mean-square error between the estimated value function $V(t,S_t;\psi)$ and the true option value function $V$, i.e., $\mathbb{E}\int_0^T| V(t,S_t;\psi)-V(t,S_t)|^2dt$. 
%Therefore, the optimal solution to problem \eqref{rho_g_dicrete} provides a well-founded estimation for the option price function $V$. 
Again, we can apply the stochastic gradient descent algorithm to minimize loss function $\text{ML}_{\Delta t}(\psi)$ and get the minimizer $\psi^*$.

In subsequent numerical experiments, we only evaluate the option price function $V_0(\cdot)$ rather than $V(\cdot, \cdot)$ because (i) numerically it is more efficient, and (ii) we are more concerned with the current option price. %And in practice, we only have the current stock price, not future stock prices. Even if we estimate $V(t, S)$, it would only provide the option price at the current time, and estimating option prices at various times $t$ is unnecessary. More importantly, the loss function used to directly estimate the function $V(t, S)$ has a more complex landscape compared to estimating $V_0(S)$. In numerical experiments, we have found that its estimates at $t=0$ have larger errors compared to $V_0(S)$.
}

\section{Numerical Results}
\label{data_test}
We test the performance of our algorithm using synthetic data generated by the following mean-reverting process of the stock price in the objective world:
\begin{equation}
	\label{equ:cir}
	dS_t=a(b-S_t)dt+{\sigma_{loc}}(S_t)  S_t dB_t.
\end{equation}
There are two reasons for considering such synthetic data. First, it is reasonable to assume such a mean-reverting process for some underlyings such as VIX, foreign exchange rates, and commodities. 
Second, the stock prices generated by a mean-reverting process are concentrated within a certain range, which enables us to efficiently compute the expectation in equation \eqref{loss_f_dl_diff}.

%{\color{red}we may draw enough (high quality) sample paths from a single trajectory generated by this mean reverting process}\footnote{In practice, the quantity of samples is limited. Here we provide an ample number of samples to assess the performance of our proposed algorithm. These samples are generated from a single path to characterize that historical price data is a trajectory of stochastic model.}, and the prices are relatively concentrated within a specific price range. The initial values $S_0$ of the samples approximately follows a Gaussian mixture, which is resulted from strongly mean reverting with a stationary limiting distribution at infinity. Thus, we can use the corresponding sample mean to estimate the expectation in formula \eqref{loss_f_dl_diff} via Bernstein theorem \citep{EUGENE}. In contrast, the samples generated by single-trend (up trend or down trend) models do not exhibit this characteristic mentioned above. The mean of initial values of samples increases (decreases) towards infinity as the sample size increases. It is difficult for the initial values $S_0$ to satisfy a specific distribution $\omega$, which means we cannot estimate its expectation by the mean of corresponding samples. %Furthermore, models where trends change over time, like the mean reverting model, are more realistic. Single-trend in long term are not commonly observed in real financial markets.

We consider two special datasets. One is with a special Cox-Ingersoll-Ross (CIR) model \citep{Cox1985}, where ${\sigma_{loc}}(S_t) = \sigma_0/ \sqrt{S_t}$, $a=0.1$, $b=1.3$, and $\sigma_0 = 0.2$.\footnote{The CIR model with these parameter values satisfies the Feller condition $2ab>\sigma_0^2$, so the stock price must be positive.} The other is with a generalized local volatility (GLV) model, where $a=3$, $b=0.98$, and the local volatility function $\sigma_{loc}(\cdot)$ is reconstructed from an implied volatility function $\sigma_{imp}(\cdot)$ through Dupire equation \citep{Dupire1994}. That is\footnote{The parameter $T^\star$ refers to the maturity of the options used to calibrate the local volatility model via the Dupire equation. Hence $T^*$ is not necessarily equal to $T$, the maturity of the options we want to price, though we take $T^*=T$ in our experiments.}
$$
\sigma_{loc}(K)=\sqrt{\frac{\sigma_{imp}^2+2 r \sigma_{imp} KT^\star \sigma_{imp}^\prime}{\left(1+K d_1 \sqrt{T^\star} \sigma_{imp}^\prime\right)^2+ \sigma_{imp} T^\star K^2 \left(\sigma_{imp}^{\prime\prime} -d_1\sigma_{imp}^{\prime 2} \sqrt{T^\star}\right)}}
$$
where $d_1=\frac{-log(K)+T^\star(r^\star+\frac{1}{2}\sigma_{imp}^2)}{\sigma_{imp}\sqrt{T^\star}}$,
$
\sigma_{imp}(K) = \left(2.681K^2 - 5.466K + 2.981 \right)\mathbf{1}_{\{0.60<K<1.33 \}} +0.667\times\mathbf{1}_{\{K\leq 0.60\}} + 0.454\times\mathbf{1}_{\{K\geq 1.33\}}
$, $\sigma_{imp}^\prime$ and $\sigma_{imp}^{\prime\prime}$ are the first- and second-order derivatives of $\sigma_{imp}$ w.r.t. $K$, respectively,  $r^\star=0.019$, and $T^\star = 0.1$. For each model, we generate a single price trajectory with initial value $S_{0}=1$ and time step size $\Delta t = 3 \times 10^{-3}$, and extract $J$ sample paths from the trajectory, each with a fixed-length time length $[t_i, t_i + T ]$, $i = 0,1 \cdots J-1$. %We take them as two separate training data sets, which are assumed to be enough. 
%We implement our algorithm on these two datasets separately.

The options we price are European-style vanilla (call/put) options with strike price $K=1.0$ and maturity $T=0.1$. We set the risk-free rate $r = 0.019$ and the continuous dividend yield $q=0$. The benchmark values are obtained by the finite difference method for the PDE models with true parameters.% and $\sigma_{loc}(\cdot)$.

To parameterize the policy function $f(\cdot)$, we use a three-layer fully connected neural network for the CIR model and a fifteen-layer residual neural network for the GLV model, where the rectified linear unit (ReLU) and 128 neurons are adopted. To parameterize the option value functions $V_0(\cdot)$, we use a six-layer fully connected neural network with ReLU and 256 neurons for call options and a four-layer fully connected neural network with LeakyReLU and 128 neurons for put options in the CIR model; we use a nine-layer fully connected neural network with LeakyReLU and 128 neurons  for call options and a thirteen-layer
residual neural network with ReLU and 128 neurons for put options in the GLV model. We employ the mini-batch stochastic gradient descent algorithm to minimize the loss functions, where the mini-batch size is chosen to be 128 and 256 for the CIR and GLV models, respectively. The above hyperparameters were selected among several candidate sets of hyperparameters based on the criterion of minimizing training losses. 

\begin{figure}[htbp]
	\centering
	%	\FIGURE
	%	{
	%		\includegraphics[scale=0.4]{./figures/policy_func_cir.eps} 
	%		\includegraphics[scale=0.4]{./figures/policy_func_local_vol.eps}
	%	}
	\begin{minipage}{0.41\textwidth}
		\centering
		\includegraphics[scale=0.37]{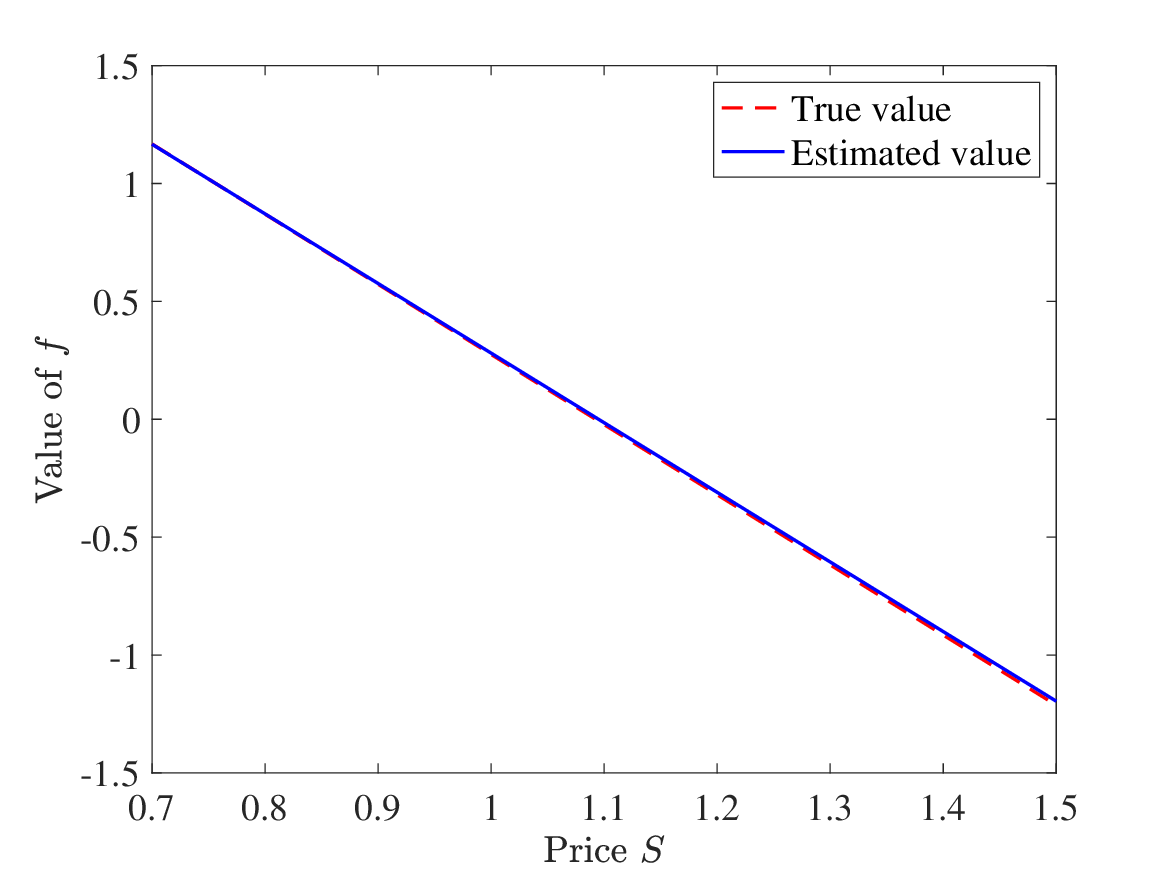} 
		{\small \ \ The CIR model}
	\end{minipage}
	\begin{minipage}{0.41\textwidth}
		\centering
		\includegraphics[scale=0.37]{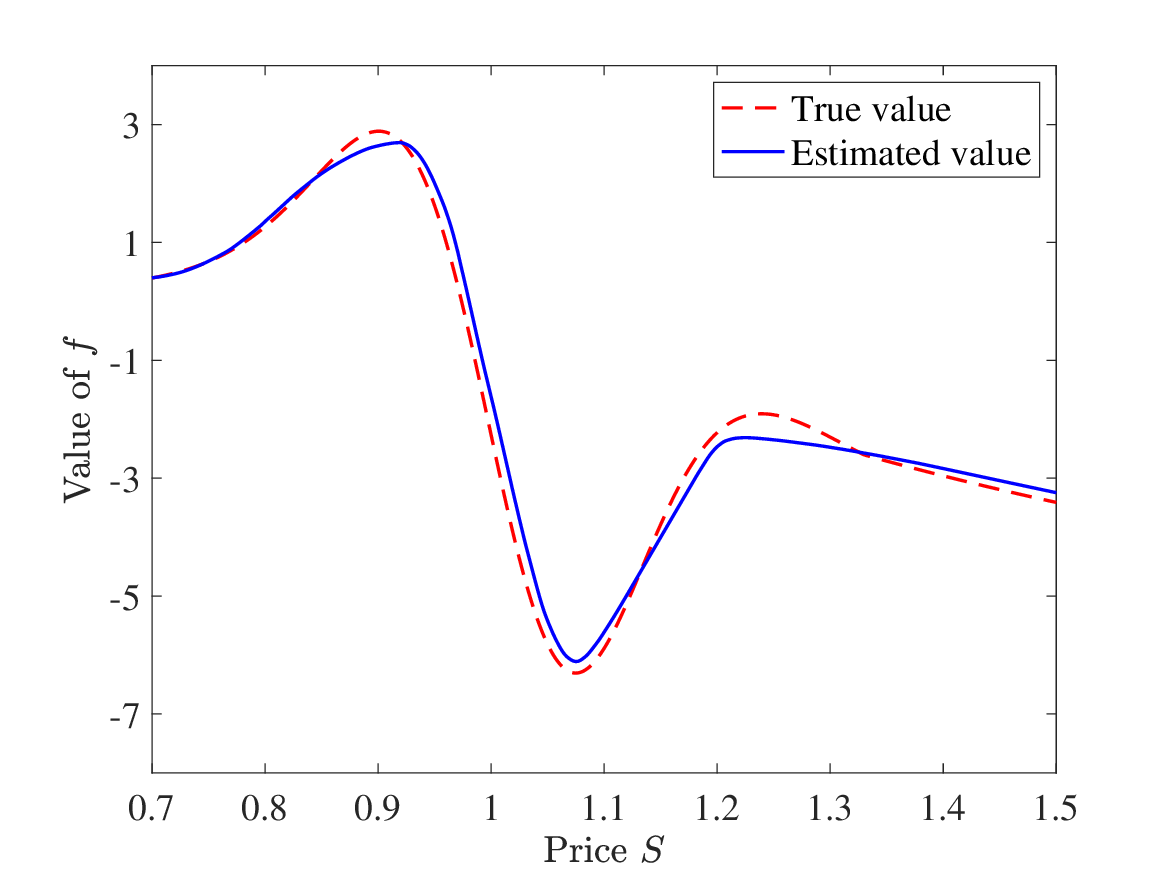} 
		{\small \ \ The GLV model}
	\end{minipage}
	\ \\
	\ \\
	\caption{{Optimal Policy Functions $f(\cdot)$ Obtained by Our Algorithm and True Parameters}
		\label{fig:policy_func}}
	%	{The left is with the CIR model, and the right is with the GLV model. }
\end{figure}

In Figure \ref{fig:policy_func}, we compare the learned optimal policy function $f(\cdot)$ obtained by our method (blue solid line) with the theoretically optimal policy function (red dashed line). The left figure is with the CIR model, while the right is with the GLV model.  It can be observed that our algorithm can capture the optimal policy function quite well. % within a certain price range.

%With estimated optimal policy function $\hat{f}(s)$, the pricing kernel $\hat{\rho}_t$ for each sample could be constructed. Next step is to learn European option value functions {\color{blue}{$V_0(s)$}}.
% Instead of using a general bivariate neural network taking both time $t$ and price $s$ as inputs, we use neural networks $NN^i(s;\varphi_i)$ taking $s$ as input to parametrize option price functions $V(t_i,s)$ for each discrete time node $t_i$. In this way, it's easier to train since the loss landscape is simpler than before. 

\begin{figure}[htbp]
	\centering
	\begin{minipage}{0.41\textwidth}
		\centering
		\includegraphics[scale=0.35]{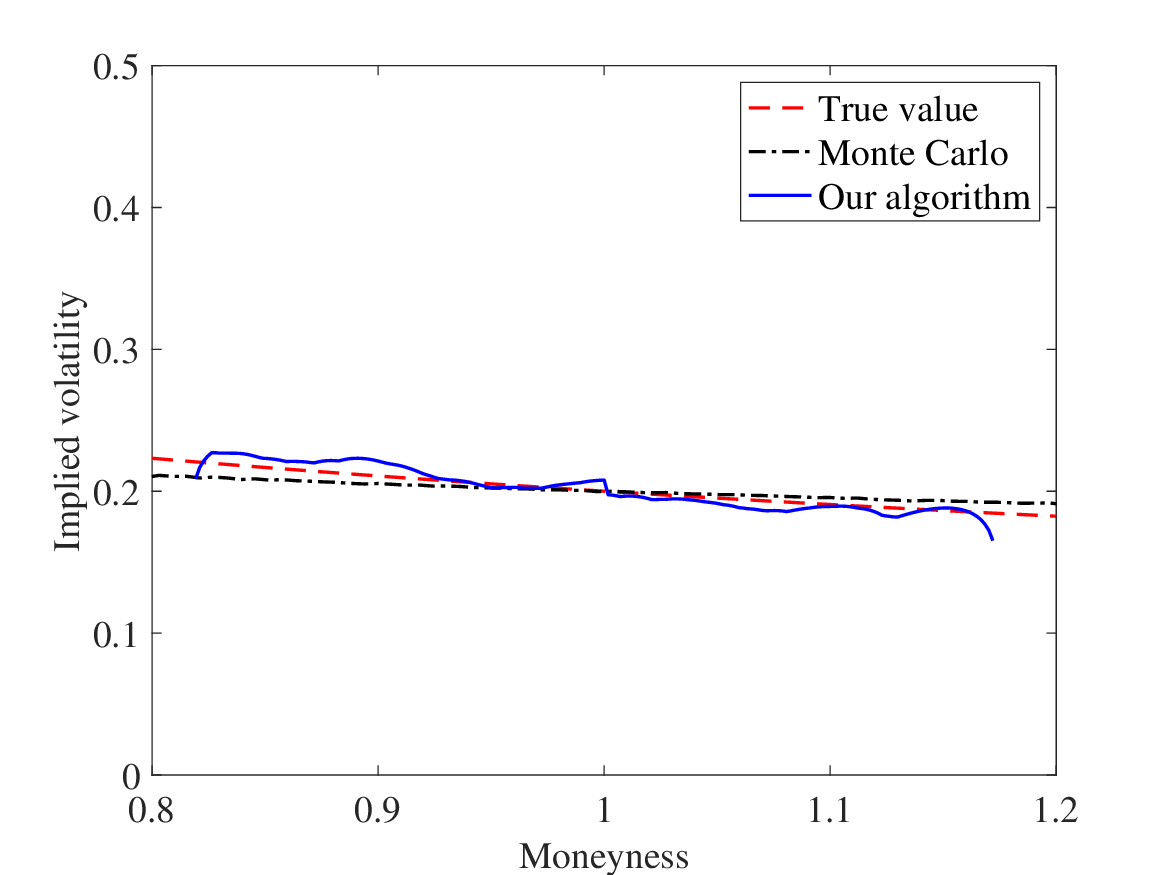} 
		{\small \ \ The CIR model}
	\end{minipage}
	\begin{minipage}{0.41\textwidth}
		\centering
		\includegraphics[scale=0.35]{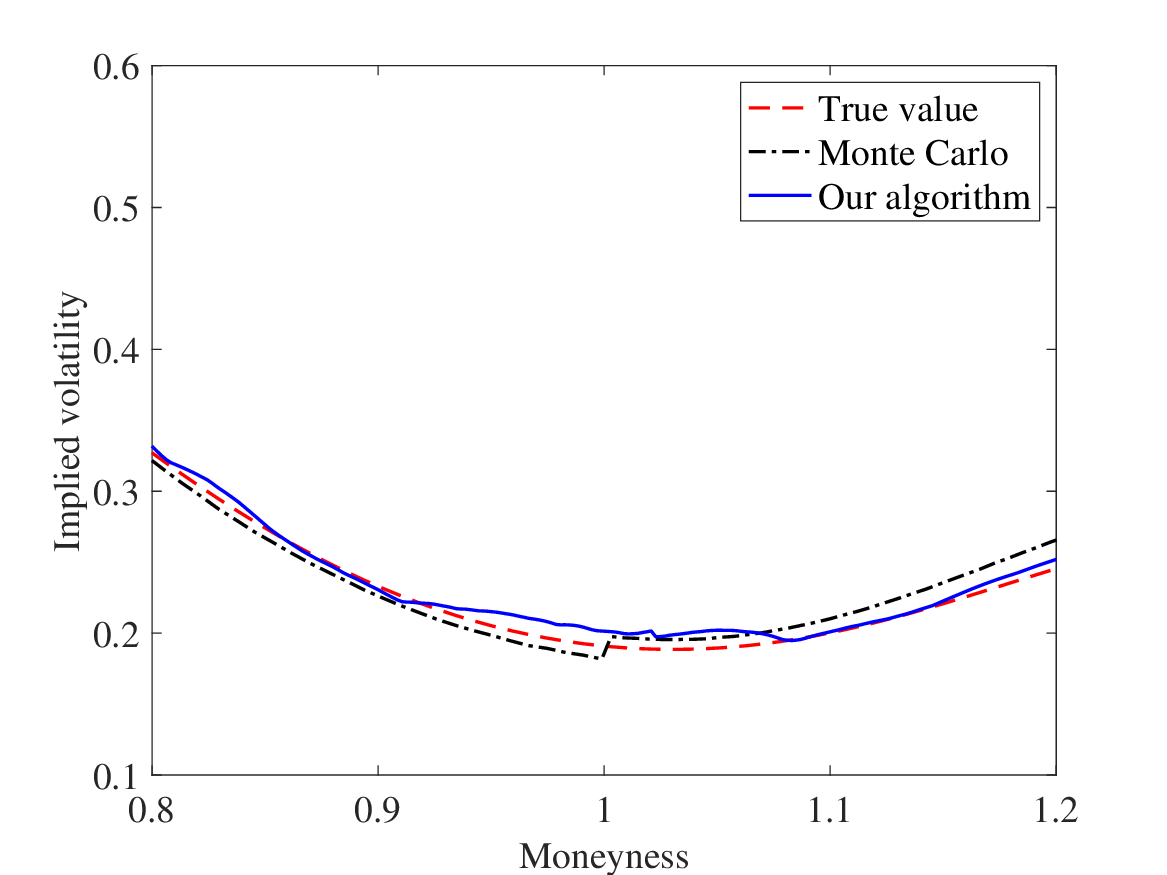} 
		{\small \ \ The GLV model}
	\end{minipage}
	\ \\
	\ \\
	\caption{{Implied Volatility Associated with the Option Prices Obtained by Different Methods}
		\label{fig:impv_algo}}
\end{figure}

Next we compare the learned option pricing functions obtained by our algorithm and the benchmark pricing functions obtained by the finite difference method for the PDE model with true parameters. To measure their difference, we transform the price functions into the implied volatility functions against the moneyness (the ratio of spot price to strike price). In particular, we focus on out-of-money options, plotting the implied volatility curves against the moneyness for call options when the moneyness is less 1, and for put options when the moneyness is greater than 1. In Figure  \ref{fig:impv_algo}, we present the benchmark implied volatility curve (red dashed line) and the learned implied volatility curve (blue solid line), indicating that our method performs very well. It is worth pointing out that our method is less accurate for deep out-of-money options. The reasoning is the following. Our method outputs a price function, which is very close to the true solution. %(see Figure \ref{fig:opt_price_func} in Appendix). %\ref{add_numerical_res}). 
However, implied volatility is very sensitive to errors %turns significant 
when option prices are lower.    %performs better near the at-the-money region and extends reasonably well to a certain range of out-of-the-money and in-the-money options.
 %This is reasonable. The farther the moneyness deviates from 1, the larger the reciprocal of Vega becomes. Even if the estimated option prices are very close to the real values, the corresponding implied volatility errors will also be significant (magnified by the reciprocal of Vega times based on option price errors).

%Meanwhile, to quantitatively assess the impact of estimation errors in the policy function $f$ on implied volatility, we employ the Monte Carlo method to approximate the option value $V_0(s)$ based on formula \eqref{price_formual}, using estimated optimal policy function $\hat{f}$ and simulation paths generated from true models. Varying $s$ and repeating the process, the values of options are obtained at different moneyness. Again, we use out-of-money option to calculate the corresponding implied volatility (dash-dotted line), as also illustrated in figure \ref{fig:impv_algo}.

Further, we investigate whether the errors in estimating the optimal policy function $f^*(\cdot)$ significantly affect the accuracy of option pricing. To do this, we assume $\sigma_{loc}(\cdot)$ is known, and for a given $S_0$, we generate random stock paths according to the real-world stock price process \eqref{equ:cir}. For each path, we can obtain $\rho_T$ through the learned optimal policy function $f^*(\cdot)$. Then we can employ the Monte-Carlo simulation to estimate the expectation as given in \eqref{price_formual}. Repeating the above procedure for different initial stock prices leads to an option price function. In Figure \ref{fig:impv_algo}, we plot the corresponding implied volatility curves using the black dash-dotted line. It can be observed that the black dash-dotted curves are close to the other two curves, implying that the estimation errors for the optimal policy function $f^*(\cdot)$ have a negligible impact on the accuracy of our pricing methodology. %This confirms the applicability of our algorithm for estimating the policy function $f$. 

\begin{figure}[htbp]
	\centering
	\begin{minipage}{0.41\textwidth}
		\centering
		\includegraphics[scale=0.37]{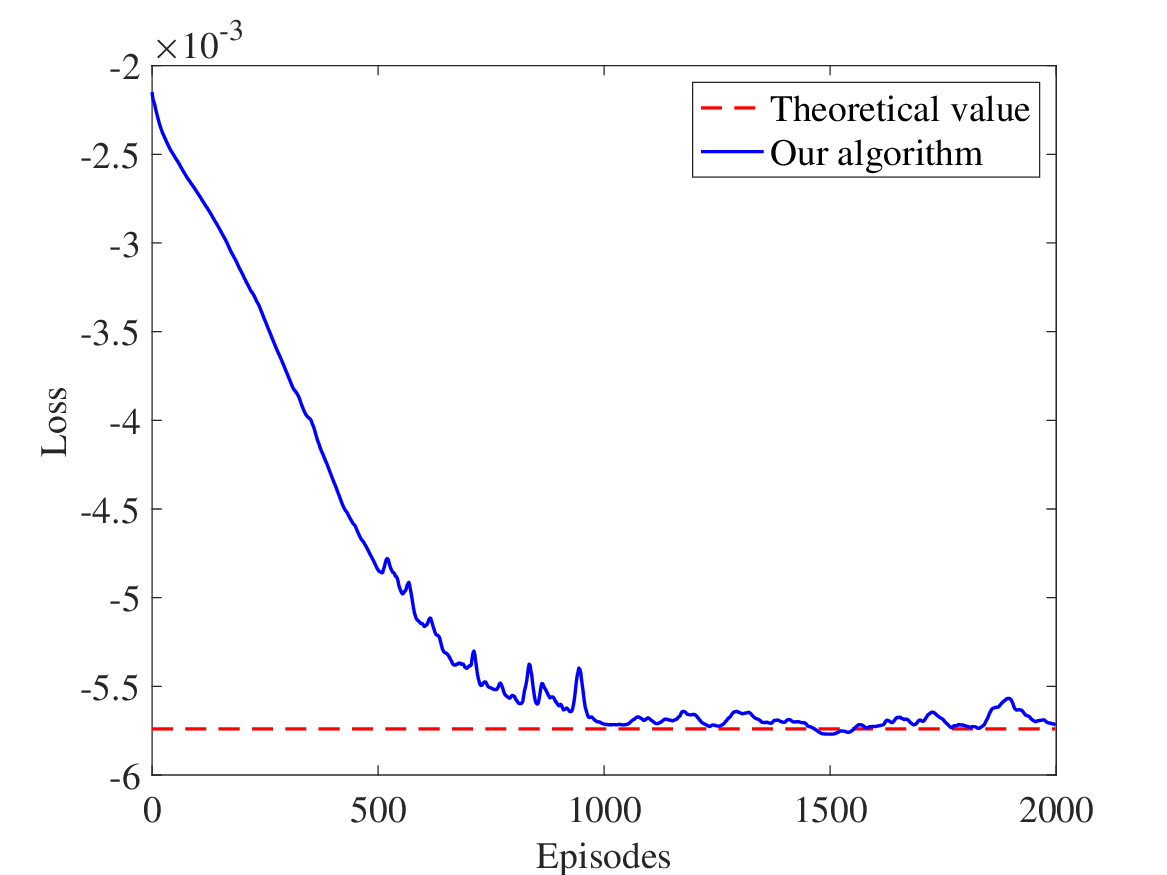} 
		{\small \ \ The CIR model}
	\end{minipage}
	\begin{minipage}{0.41\textwidth}
		\centering
		\includegraphics[scale=0.37]{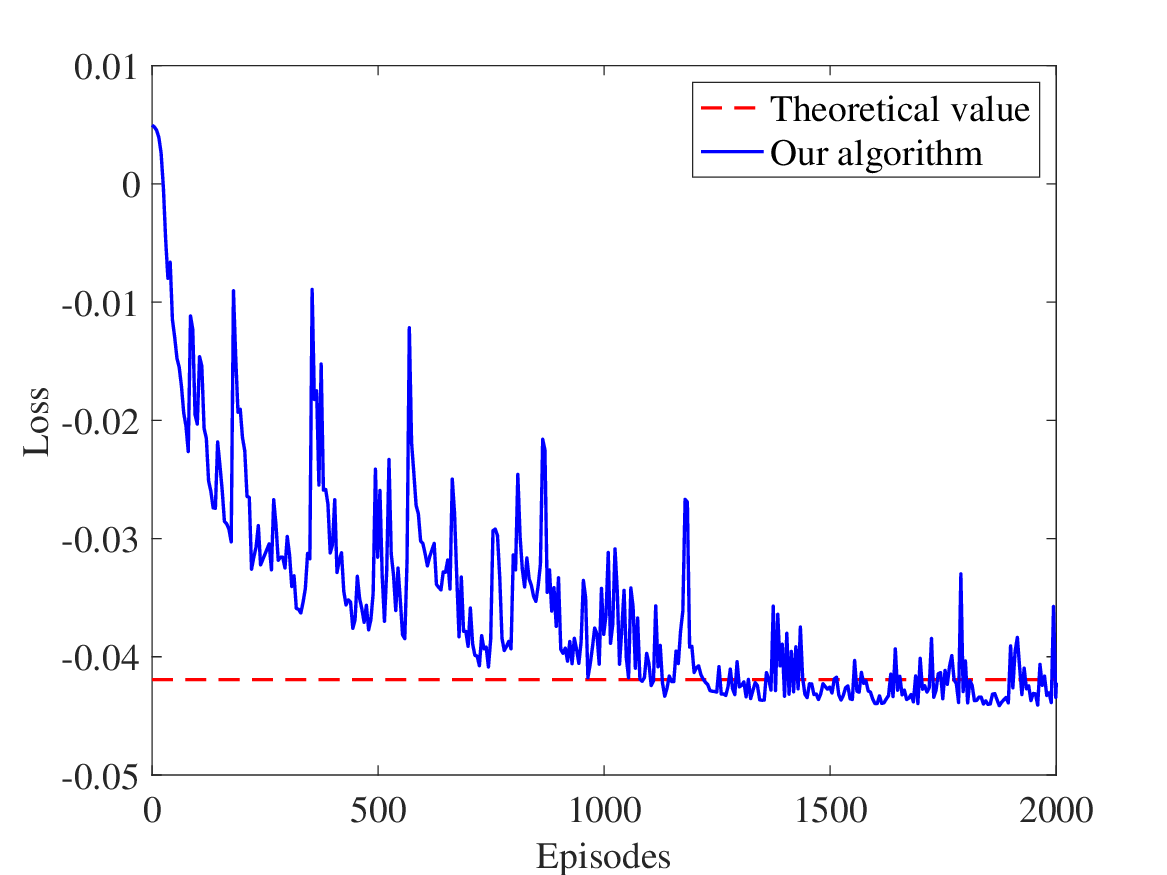} 
		{\small \ \ The GLV model}
	\end{minipage}
	\ \\
	\ \\
	\caption{{Loss Curves against the Number of Episodes when Learning the Optimal Policy Function}
		\label{fig:log_return_episodes}}
\end{figure}

Lastly, we investigate the convergence of our algorithms against the number of episodes. For illustration, we consider the case for learning the optimal policy function. 
We take additional 25,600 sample paths as the validation dataset. For each episode in the training procedure, we also compute the loss function over the validation dataset, which we call the loss curve. In Figure \ref{fig:log_return_episodes}, 
we plot the loss curve (the blue solid line) against the number of episodes for the CIR model (left panel) and GLV model (right panel), respectively. For comparison, we also plot the theoretical value (the red dashed line) that the loss curve converges to.\footnote{The theoretical value is nothing but the negative value function associated with the portfolio optimization problem.} %, equal to expected terminal utilities are around $5.74\times 10^{-3}$ and $4.19\times 10^{-2}$, respectively, as shown by red dashed lines.\footnote{The expectations are estimated by the mean of terminal utilities in validation set.} 
Notably, after approximately 1,000 episodes, the loss curve converges towards the theoretical value. %This result also reflects a low level of sample complexity of our algorithm for learning the optimal policy function because it requires more than 128,000 sample paths. This limitation poses a risk that the quantity of samples necessitated by our algorithm may exceed the amount  of relevant available historical data. Such challenges related to sample efficiency are not exclusive to our algorithm but are commonly encountered in the field of learning trading strategies through deep learning or reinforcement learning techniques \citep{Hambly2021}.

\section{Future Research Directions}
For potential future research directions along this line, we may enhance the deep learning algorithm used in our approach so that the amount of data required is close to reality.  Besides, %this paper focuses on a complete market in which there is a unique pricing kernel. 
it is worth extending our pricing methodology to an incomplete market where infinitely many pricing kernels exist. 

%The results validated that our algorithm can effectively estimate option prices within regions of concentrated underlying price data distribution. In the meanwhile, it reveals that the historical price of the underlying asset inherently captures its volatility. %It is not necessary to calibrate its volatility from the price of its derivatives. 
%This provides an alternative perspective for option pricing.

\begin{appendices}
\section{Appendix}
\paragraph{Proof of Theorem \ref{theorem_log_util}.}
\label{proof_theorem_log_util}
%	\proof{Proof.}
		It is easy to see that the optimal terminal wealth for problem \eqref{log_util} is the optimal solution for
		\begin{equation}
			\max \mathbb{E}[\ln (\xi)] \quad \text { s.t. } \mathbb{E}\left[\xi \rho_T\right]=1,~\xi>0 .
		\end{equation}
		The optimal terminal wealth is $\xi=\rho_T^{-1}$. Hence the wealth process is$X_t=\mathbb{E}\left[\xi \rho_T / \rho_t \mid \mathcal{F}_t\right]=\rho_t^{-1} $.
		Since $d \rho_t=\rho_t\left(-r d t-\theta_t d B_t \right)$
		with $\theta_t\in \mathbb{R}$ satisfying $\theta_t=\frac{\mu\left(S_t\right)-r}{\sigma(S_t)},$
		we have
		$$
		d X_t  = d(\rho_t^{-1})= X_t\left[r+\frac{\theta_t}{ \sigma(S_t)}(\mu(S_t)-r) \right] d t+X_t \theta_td B_t .
		$$
		which means $\pi_t^*=X_t\frac{\theta_t}{ \sigma(S_t)}$. Notice that $\pi_t^*$ is a feedback policy and observable with respect to $\mathcal{F}_t^{S, r}$.
		\Halmos
	\endproof

\paragraph{Proof of Proposition \ref{prop_loss_same_min}.}
\label{proof_prop_loss_same_min}
%\proof{Proof.}
According to Theorem \ref{theorem_log_util}, for each $s_0$, $f^*(\cdot ) = \frac{\mu(\cdot)-r}{\sigma^2(\cdot)}$ is the unique maximizer of $\hat{L}(f\mid s_0)$ and is independent of $s_0$. %Hence $\hat{L}(f\mid s_0)$ has the same unique maximizer $f^*$ whenever $s_0$ is. 
It is clear that $f^*$ also maximizes $L(f) = \mathbb{E}_{s_0\sim\omega}[\hat{L}(f\mid s_0)]$.

%Next we prove that $f^*$ is the unique maximizer of $L(f)$. 
Suppose $L(f)$ has another maximizer $f_0 \neq f^*$, then $\hat{L}(f_0\mid s_0) = \hat{L}(f^*\mid s_0)$ for any $s_0$. Hence we have
$$L(f_0) = \mathbb{E}_{s_0\sim\omega}[\hat{L}(f_0\mid s_0)] = \mathbb{E}_{s_0\sim\omega}[\hat{L}(f^*\mid s_0)] = L(f^*).$$
Hence we have $\hat{L}(f_0\mid s_0)=\hat{L}(f^*\mid s_0)$ for almost all $s_0$, 
which contradicts the fact that $f^*$ is the unique maximizer of $\hat{L}(f\mid s_0)$.
\Halmos
\endproof
%\paragraph{Figure \ref{fig:opt_price_func}.} It presents the learned and true price functions of call and put options for the CIR model.
%\label{add_numerical_res}
%	\begin{figure}[htbp]
%	\centering
%%	\begin{minipage}{0.41\textwidth}
%%		\centering
%		\includegraphics[scale=0.35]{./figures/option_price_cir.eps} 
%%		{\small \ \ The CIR model}
%%	\end{minipage}
%%	\begin{minipage}{0.41\textwidth}
%%		\centering
%%		\includegraphics[scale=0.37]{./figures/option_price_glv.eps} 
%%		{\small \ \ The GLV model}
%%	\end{minipage}
%	\caption{{European Vanilla Call/Put Option Price Functions Learned by Our Algorithm for the CIR Model}
%		\label{fig:opt_price_func}}
%\end{figure}

\end{appendices}
\SingleSpacedXI

\bibliographystyle{informs2014}
\bibliography{myrefs}

\end{document}